\documentclass[aps, prmaterials, twocolumn, longbibliography, 10pt]{revtex4-2}

\usepackage[a4paper, total={7in, 10.3in}]{geometry}
\usepackage[utf8]{inputenc}
\usepackage[english]{babel}
\usepackage[font=small,labelfont=bf, justification=raggedright,singlelinecheck=false]{caption} 
\usepackage{amsfonts, amsmath, amsthm, amssymb} 
\usepackage{braket}
\usepackage{graphicx}
\usepackage{footnote}
\usepackage{tabularx}
\usepackage[plainpages=false,pagebackref=false,pdftex]{hyperref}

\hypersetup{hyperindex,colorlinks=true,breaklinks=true,linkcolor=black,citecolor=black, urlcolor=black}

\begin{document}
    
\title{How to choose efficiently the size of the Bethe-Salpeter Equation Hamiltonian for accurate exciton calculations on supercells}

\author{Rafael R. Del Grande}
\email{rdelgrande@ucmerced.edu}
\author{David A. Strubbe}
\email{dstrubbe@ucmerced.edu}
\affiliation{Department of Physics, University of California, Merced, California, USA}

\date{\today}

\begin{abstract}

The Bethe-Salpeter Equation (BSE) is the workhorse method to study excitons in materials. The BSE Hamiltonian size, which depends on how many valence-to-conduction band transitions are considered, needs to be chosen to be sufficiently large to converge excitons' energies and wavefunctions but should be minimized to make calculations tractable, as BSE calculations are expensive and scale with the number of atoms as $\mathcal{O}(N_{\rm{atoms}}^6)$. In particular, in the case of supercell (SC) calculations composed of $N_{\rm{rep}}$ replicas of a primitive cell (PC), a natural choice to build this BSE Hamiltonian is to include all transitions derived from PC calculations by zone folding. However, this leads to a very large BSE Hamiltonian, as the number of matrix elements in it is $(N_k N_c N_v)^2$, where $N_k$ is the number of $k$-points and $N_{c(v)}$ is the number of conduction (valence) states used. When creating a SC, the number of $k$-points decreases by a factor $N_{\rm{rep}}$ but both the number of conduction and valence states increase by the same factor, therefore the number of matrix elements in the BSE Hamiltonian  increases by a factor $N_{\rm{rep}}^2$, making exactly corresponding calculations prohibitive. Here, we provide a workflow to decide how many transitions are necessary to achieve comparable results, based on only PC results. With our method, we show that to converge the first exciton binding energy of a LiF SC composed of 64 PCs, to an energy tolerance of 0.15 eV, we only need 12\% of the valence-to-conduction matrix elements that result from zone folding with a minimal set of bands. As an example, we use the number of bands from our method to obtain the absorption spectrum of LiF with a V$_k$-like defect. The procedure in our work helps in evaluating excitonic properties in large SC calculations, such as moir\'e patterns, charge density waves, defects, self-trapped excitons, polarons, interfaces, and other kinds of disorder and can be used within any code that builds the BSE Hamiltonian.

\end{abstract}

\maketitle

\section{Introduction}
\label{sec:introcution}

    \textit{Ab initio} calculations of crystals are widely used in the materials science community. Most of those techniques use crystal periodicity and Bloch's theorem to understand the electronic properties of materials. However, there are many cases of interest where the translational symmetry is broken, such as moir\'e patterns of 2D materials \cite{Naik2022Nature, Susarla2022}, charge density waves \cite{CDW}, defects \cite{DelBen2019CompPhysComm}, alloys \cite{SQS}, amorphous materials \cite{AnhPham2013, Strubbe2015PRB, Winkler2017}, self-trapped excitons \cite{IsmailBeigi2005PRL, delgrande2025arxiv}, frozen phonon calculations \cite{2020Cook}, polarons \cite{Franchini2021NatRevMat}, surfaces \cite{AnhPham2013, WangPRB2003}, thermal disorder \cite{Zacharias}, and cases of structural and magnetic symmetry-breaking \cite{ZhaoZunger,WangZunger}. In this kind of situation, it is common to use a supercell (SC) approximation \cite{Freysoldt2014RevModPhys} where the SC size must be large enough to converge the desired properties \cite{JihyePRB2005, Castleton2009}, such as electronic energy levels or phonon frequencies. The SC consists of $N_{\rm{rep}}$ replicas of one Primitive Cell (PC), and the maximum size that can be used is limited in practice by the available computational resources. Schemes to reduce the computational expense are often necessary. In particular, if the atoms are only slightly moved from their original positions due to a symmetry breaking, such as local strains and defects, then SC properties are still similar to PC properties, which can be used to construct an approximation for Bethe-Salpeter Equation (BSE) calculations.


DFT studies using the SC approximation are already common \cite{Freysoldt2014RevModPhys}, but GW and BSE calculations are less common due to their high computational cost \cite{DelBen2019CompPhysComm, AbtewPRL2011, barker2022arxiv, Lischner2012PRL, MaPRL2010}. The GW method \cite{Hybertsen1986} is known to have good performance in reproducing the experimental bandgaps of semiconductors, while BSE \cite{Rohlfing2000, Onida2002, Blase2020, Leng2016, Strinati1988} is successful in studying excitonic effects. GW/BSE calculations are performed on top of DFT results and are, in general, more computationally demanding than DFT calculations, and their convergence is harder to achieve. Typical plane-wave DFT calculations scale with $\mathcal{O}(N_{\rm{atoms}}^3)$ \cite{Payne1992RevModPhys}, while GW/BSE calculations scale up to $\mathcal{O}(N_{\rm{atoms}}^5)-\mathcal{O}(N_{\rm{atoms}}^6)$ \cite{Deslippe2012}, where $N_{\rm{atoms}}$ is the total number of atoms. There have been advances in parallel efficiency \cite{DelBen2019CompPhysComm} and various approximations and numerical schemes that can be used to handle bigger SCs.
To compute the screened Coulomb interaction (the $W$ in the GW approximation), one can make use of analytical models for the dielectric function \cite{Forde2023NanoLett, Trolle2017SciRep}. The stochastic pseudobands method speeds up GW calculations from $\mathcal{O}(N^4)$ to $\mathcal{O}(N^2)$ \cite{AltmanPRL2024}. Scissor-operators are also a valid option in the case where the GW corrections follow approximately a linear relation with DFT energy levels \cite{Peelaers2011, Deslippe2012}. Subsampling schemes deal with $\mathbf{G}$-vectors perpendicular to the layers of 2D systems, which greatly reduces the density of $\mathbf{k}$-point sampling the first Brillouin Zone (BZ) \cite{Jornada2017PRB} for both GW and BSE calculations. One can also expand wavefunctions of SC on the basis of wavefunctions of PC, and then calculate SC kernel matrix elements in terms of PC kernel matrix elements \cite{Strubbe2012, Naik2022Nature, Susarla2022}. Dai \textit{et al.} have developed a variational method that combines finite-momentum excitons and phonons to study polaronic excitons without the necessity of SCs \cite{Dai2024PRB, Dai2024PRL}. Tight-binding \cite{Zanfrognini2023, Cho2019, Bieniek2022, Dias2023} and machine learning \cite{Dong2021, venturella2024} approaches to GW/BSE also provide a significant speed-up but need to be parametrized very accurately. Despite this work, new theories and approaches are still necessary to improve the performance of GW/BSE calculations in bigger SCs. 

In this paper, we focus on the BSE calculations, and more specifically on the kernel matrix element calculations, which take into account the electron-hole interaction and are typically the most computationally demanding step of our SC GW/BSE calculations \cite{Deslippe2012}. The kernel calculation scales as $\mathcal{O}(N_{\rm{atoms}}^5)$ and the absorption step, in which the BSE Hamiltonian matrix is diagonalized, scales as $\mathcal{O}(N_{\rm{atoms}}^6)$, but typically the prefactor makes this step considerably less expensive \cite{Deslippe2012}. The main convergence issue in this kind of calculation is the number of conduction and valence bands to include in the BSE Hamiltonian. The number of bands can be very large, leading to a large number of node-hours and amount of RAM needed to perform the calculations, as well as a large amount of disk space needed to store wavefunctions and GW/BSE results, making BSE calculations on SCs computationally very challenging. We provide an efficient way of choosing the number of valence and conduction for SC based on the results of PC calculations. Within our scheme, we built a BSE Hamiltonian with only 12\% of the number of matrix elements that would result from zone folding of a minimal set of PC conduction and valence states.  We achieved kernel calculations 8.4 times faster with errors of only 150 meV for the binding energies of the first exciton in SC calculations compared to PC results for rocksalt LiF. Herein we use the BerkeleyGW code \cite{Deslippe2012}, but this method can be easily extended to other codes that build and diagonalize the BSE Hamiltonian \cite{yambo_code, west_code, abinit_code, exciting_code}.

We chose to study LiF as it is an insulator with strong excitonic effects that are well studied theoretically \cite{Rohlfing2000}. Due to the strong electron-phonon interaction, polaronic effects are strong as well \cite{SioPRL2019, SioPRB2019}, and it presents self-trapped excitons with polaronic character \cite{Dai2024PRB, Dai2024PRL,delgrande2025arxiv} and color centers \cite{Baldacchini2001, ShlugerPRB1991, SongPRB1992}. Therefore LiF is an excellent choice for testing new methodologies for BSE supercell calculations, although our method does not make any assumptions about the material, so it can be applied to any material where excitonic effects are present, such as 2D materials \cite{Qiu2013, Naik2022Nature, Susarla2022, Zanfrognini2023, DadkhahPRB2024}, carbon nanotubes \cite{tese_rafael}, etc. 

In this work we use the term ``exciton'' for all eigenvectors of the BSE Hamiltonian, regardless if their energies are below or above the bandgap, as often done in literature on BSE \cite{Deslippe2012}. Usually one reserves the term ``exciton'' for transitions below the bandgap, as experimentally excitons' signature are sharp peaks below the continuum in optical absorption experiments, which indicates they are bound excitons, and their binding energy is defined as the difference between the excitation energy and the bandgap. From a more analytical point of view, the binding energy may be defined by the action of the kernel operator \cite{Rohlfing2000} ($E_b=\langle A|K^{\rm{eh}}|A\rangle$), and even transitions with energy above the bandgap can be classified as ``bound'' excitons if they are composed of higher-energy transitions. For example, recently, an experimental paper studied the dark excitons in CrI$_3$ with energies above the bandgap \cite{He2025PRX}.

The paper is organized in the following way: in Section \ref{sec:theory} we discuss the zone-folding aspects to be considered in BSE calculations and our approach, in Section \ref{sec:computational_details} we provide computational details of our calculations, in Section \ref{sec:results} we show our approach applied to LiF SC calculations, in Section \ref{sec:Vk_center_results} we calculate the absorption spectra of LiF with a defect similar to a V$_k$ center , and in Section \ref{sec:conclusions} we present our conclusions.

\section{Theory}
\label{sec:theory}

\subsection{Consideration about $k$ folding}

Since the concepts of $k$-point sampling and total number of $\bf{G}$-vectors are common for both DFT and GW/BSE calculations, we start discussing them in the case of DFT. Let's consider a crystal with lattice vectors $\boldsymbol{a_1}, \boldsymbol{a_2}$ and $\boldsymbol{a_3}$ and reciprocal lattice vectors $\boldsymbol{b_1}, \boldsymbol{b_2}$ and $\boldsymbol{b_3}$ which obey $ \boldsymbol{a_i} \cdot \boldsymbol{b_j} = 2 \pi \delta_{ij}$ \cite{Martin_Book}. If a SC has lattice vectors $N_1 \boldsymbol{a_1}$, $N_2 \boldsymbol{a_2}$, and $N_3 \boldsymbol{a_3}$, with $N_i \geq 1$ being a natural number, then the primitive reciprocal lattice vectors are given by $\boldsymbol{b_1} / N_1$, $\boldsymbol{b_2} / N_2$, $\boldsymbol{b_3} / N_3$. Therefore, the SC volume increases by a factor $N_{\rm{rep}} = N_1 N_2 N_3$, while the first BZ reciprocal volume decreases by this same factor. In the limit of very large SC, just the $\Gamma$ point is sampled \cite{DelBen2019CompPhysComm}. The results of a SC calculation will be identical to results of a PC calculation when the number $N_k^{\rm SC}$ of $k$-points in a regular grid for the SC relates to the number $N_k^{\rm PC}$ of $k$-points in a regular grid of a PC as $N_k^{\rm SC} = N_k^{\rm PC} / N_{\rm{rep}}$. For example, a PC calculation of a crystal with $8 \times 8 \times 8$ $k$-grid corresponds to a $4 \times 2 \times 1$ $k$-grid in a $2 \times 4 \times 8$ SC. Given that $N_k^{\rm SC}$ and $N_k^{\rm PC}$ are integers, this condition cannot always be met, especially if one considers more general SCs in which the lattice vectors are not multiples but linear combinations of the PC ones, e.g. for the conventional and primitive cells of diamond-structure silicon. In such cases, one can make SC results comparable to PC results by making the density of $k$-points in both BZs similar.

Plane-wave based DFT codes expand periodic quantities as $f(r) = \sum_{\bf{G}} f_{\bf{G}} e^{i\bf{G}\cdot r}$ \cite{Martin_Book}. Those expansions are done for $\bf{G}$-vectors that obey $\hbar^2|\mathbf{G}+{k}|^2/2m_e \leq E_{\rm{cutoff}}$, where $E_{\rm{cutoff}}$ is the cutoff energy. As the $\bf{G}$-vectors are linear combinations of the primitive reciprocal lattice vectors, the density of $\bf{G}$-vectors in reciprocal space increases by a factor $N_{\rm{rep}}$, so SC calculations will have more $\bf{G}$-vectors for a fixed $E_{\rm{cutoff}}$. Plane-wave DFT codes iteratively diagonalize $N_k^{\rm PC}$ Kohn-Sham (KS) Hamiltonians, with a time that scales each as $\mathcal{O}(N_{\bf{G}}^2 \log{N_{\bf{G}}})$. When dealing with a SC, the number of KS Hamiltonians decreases by a factor $N_{\rm{rep}}$ and each diagonalization will be $\sim N^2_{\rm{rep}}$ times more expensive, so the whole SC calculation will be $\sim N_{\rm{rep}}^{-1} \times N_{\rm{rep}}^{2} = N_{\rm{rep}}$ times more expensive. This increase can also be thought of as coming from the fact that the PC Hamiltonian is block-diagonal in k-point but the SC Hamiltonian loses this structure when the k-points are folded together. However, we will see that the effect is more dramatic in BSE.

DFT calculations are known to underestimate the bandgap of semiconductors and insulators. One solution to this is many-body perturbation theory. The GW approximation has had success in reproducing the electronic bandgap of a wide variety of materials \cite{Hybertsen1986, Deslippe2012,louiebook06}. To take into account excitonic effects, one can solve the Bethe-Salpeter Equation (BSE) \cite{Rohlfing2000, Deslippe2012, Strinati1988}, given by $H^{\rm{BSE}}|A\rangle =\Omega_A|A\rangle$, where $|A\rangle$ is the exciton eigenvector, $\Omega_A$ is the exciton energy and $H^{\rm{BSE}}$ is the BSE Hamiltonian. Expressed in the basis of transitions $v\mathbf{k} \to c\mathbf{k}$ (represented by the ket $|cv\mathbf{k}\rangle$), it is given by
%
\begin{equation}
    \begin{split}
        \langle cv\mathbf{k} | H^{\rm{BSE}} | c'v'\mathbf{k'} \rangle &= (E^{\rm{QP}}_{c\mathbf{k}} - E^{\rm{QP}}_{v\mathbf{k}}) \delta_{\mathbf{k},\mathbf{k'}} \delta_{cc'} \delta_{vv'} \\
        &\quad + \langle cv\mathbf{k} | K^{\rm eh} | c'v'\mathbf{k}'\rangle  
    \end{split}
    \label{eq:BSE}
\end{equation}
where the first term on the right side is the independent particle transition energy, diagonal in the $|cv\mathbf{k}\rangle$ basis, and the second term is the kernel that takes into account the electron-hole interaction. $E^{\rm{QP}}_{c(v)\mathbf{k}}$ is the quasi-particle energy for the conduction $c$ (valence $v$) state at $\mathbf{k}$, from GW calculations. The exciton wavefunction projection on the $v\mathbf{k} \to c\mathbf{k}$ basis is given by $A_{cv\mathbf{k}} = \langle cv\mathbf{k} | A \rangle$. 

One chooses the number of conduction ($N_c$) and valence ($N_v$) bands to be large enough to converge absorption peak energies of interest. In the case of small unit cells, bandwidths may be as large as a few eV, so the energy of first absorption peaks may converge with only a few bands. In the case of SCs, bandwidths are smaller due to zone folding, so more bands are necessary. This is similar to isolated  molecules where a few hundred conduction states are necessary to converge exciton energies \cite{Qiu2021PRB}, as the energy levels are dispersionless in a vacuum supercell. If the SC is $N_{\rm{rep}}$ times the size of a PC, then the SC would demand $N_{\rm{rep}}$ times the number of valence and conduction bands of PC calculations and $1/N_{\rm{rep}}$ the number of $k$-points. The total number of kernel matrix elements in PC calculation is $(N_c N_v N_k)^2$, so for the SC case, it will be $N_{\rm{rep}}^2$ times the number of matrix elements of the PC case. In addition to that, direct kernel matrix elements demand double summations over $\bf{G}$-vectors \cite{Deslippe2012}, and the total number of $\bf{G}$-vectors increases by a factor of $N_{\rm{rep}}$. Therefore, each kernel matrix element becomes much more computationally demanding, by a factor of $N_{\rm{rep}}^2$. The total expense with both the increased number of matrix elements, and expense per matrix element, scales by a factor of $N_{\rm{rep}}^4$. To illustrate this concept, in Table \ref{tab:timing_PC_vs_SC} we show the time spent on GW/BSE calculations for both PC and SC cases. The computational cost of kernel calculations increases greatly with the increase of the number of bands to build the BSE Hamiltonian and is the most demanding part of our SC GW/BSE workflow \cite{Deslippe2012}. By contrast, the absorption (diagonalization) step scales as the square of the matrix dimension, equal to the number of matrix elements, thus only as $\mathcal{O}(N_{\rm rep}^2)$, so the kernel calculation dominates.

Another issue in these calculations is the RAM usage. In particular, BerkeleyGW \cite{Deslippe2012} with standard memory options stores the matrices $M_{n,n'}(\mathbf{k}, \mathbf{q}, \mathbf{G})=\langle n,\mathbf{k}+\mathbf{G}|e^{i(\mathbf{k}+\mathbf{G})\mathbf{r}}|n',\mathbf{k}\rangle$ in RAM to compute kernel matrix elements, which require a memory quantity proportional to $(N_c+N_v)^2N_{\mathbf{k}}^2N_{\mathbf{G}}$. When moving to the supercell with all the bands from zone folding, the RAM requirements to compute each kernel matrix element scale by a factor of $N_{\rm{rep}}$. Our method, by reducing the number of bands used, also reduces the RAM requirements and makes the calculations more tractable.

\begin{table*}
    \centering
    \begin{tabular}{|c|r|r|r|c|}
    \hline
         Calculation Step & Time (s) & Nodes  & Node-hours & Notes \\
         \hline
         \hline
         PC ($n_k = 64$)  & & & & \\
         Parabands   & 11.3  & 1 & 0.003  & 157 states generated\\ 
         Epsilon     & 38.8  & 1 & 0.011 & \\
         Sigma       & 676.7 & 1 & 0.188 & $n_{\rm bands} = 22$ \\
         Kernel      & 15.0  & 1 & 0.004 & $N_v = 5, N_c = 10$ \\
         Absorption  & 36.8  & 1 & 0.010 & $N_v = 5, N_c = 10$ \\
         \hline
         SC ($n_k = 1$)  & & & & \\
         Parabands& 230.7 & 60 & 3.8  & 174338 states generated\\ 
         Epsilon    & 24.6  & 10 & 0.07 & \\
         Sigma      & 871.4 & 10 & 2.4 & $n_{\rm bands} = 256$ \\
         Kernel     & 8570.3  & 200 & 476.1 & $N_v = 137, N_c = 31 $ \\
                    & 3805.7  & 40  & 42.3 & $N_v = 61, N_c = 20 $ \\ 
                    & 275.5   & 40  & 3.1 & $N_v = 31, N_c = 11 $ \\ 
                    & 17.2    & 40  & 0.2 & $N_v = 13, N_c = 5  $ \\
         Absorption & 63.35   & 2   & 0.035 & $N_v = 137, N_c = 31 $ \\
        \hline
        SC naive estimations ($n_k = 1$)  & & & & \\
         Kernel     & $\sim 2\times 10^7 $   & 200  & $\sim 1.1\times 10^6 $ & $N_v = 320, N_c = 640 $ \\
                    & $\sim 7.2\times 10^4$  & 200  & 3986  & $N_v = 192, N_c = 64 $ \\ 
         Absorption & $\sim 1.5\times 10^5$  & 2    & 81.8 & $N_v = 320, N_c = 640 $ \\
                    &  530                   & 2    & 0.3  & $N_v = 192, N_c = 64 $ \\
         \hline
    \end{tabular}
    \caption{Timing of LiF GW/BSE calculations on Frontera at the Texas Advanced Computing Center \cite{FronteraCluster} using CPUs for LiF PC and SC (4$\times$4$\times$4). Each CPU node is a Intel 8280 ``Cascade Lake'' with 56 cores. Each row corresponds to a different step of the GW/BSE workflow in the BerkeleyGW code labeled by the name of the executable \cite{Deslippe2012}. Parabands is the executable used to generate empty states to be used in GW calculations, Epsilon calculates the dielectric matrix, Sigma calculates GW corrections, Kernel calculates electron-hole interactions, and Absorption uses data from previous steps to interpolate, build, and diagonalize the BSE Hamiltonian, obtaining the absorption spectra. More computational details can be found in Section \ref{sec:computational_details}). The most demanding step for SC is the kernel calculation. GW/BSE calculations were performed with BerkeleyGW version 4.0, compiled with Intel compilers and OpenMPI. The lowest section corresponds to ``naive'' choices of the number of bands based only on zone folding from $N_v =5, N_c=10$ and $N_v=3, N_c = 1$ bands of PC calculations. These timings are extrapolated from the SC calculations, proportional to $\left( N_c N_v N_k \right)^2$. For GW calculations $n_{\rm{bands}}$ is the number of calculated bands.}
    \label{tab:timing_PC_vs_SC}
\end{table*}

One can try to reduce the size of the BSE Hamiltonian by performing a convergence study, although this process may be very expensive, especially for larger SCs. In our method, we estimate the proper number of valence and conduction bands based only on the results of PC calculations. 

\subsection{Exciton coefficients over the Brillouin Zone}

For an exciton $A$, we define the Weighted Average Single-Particle Energies (WASPEs) as 
\begin{align}
    \bar{E}^{\rm QP}_{A,\rm{cond}} = \sum_{cv\mathbf{k}} |A_{cv \mathbf{k}}|^2 E^{\rm QP}_{c \mathbf{k}} \nonumber \\
    \bar{E}^{\rm QP}_{A,\rm{val}} = \sum_{cv\mathbf{k}} |A_{cv \mathbf{k}}|^2 E^{\rm QP}_{v \mathbf{k}},    \label{eq:energy_window_eq}
\end{align}
%
 which define a scale for the relevant transitions that compose this exciton. We highlight that the quantity
 \begin{align}
 \langle A |H^{\rm BSE,QP}| A \rangle =
 \sum_{cv{\bf k}} \left| A_{cv{\bf k}} \right|^2 (E^{\rm{QP}}_{c\mathbf{k}} - E^{\rm{QP}}_{v\mathbf{k}}) \delta_{\mathbf{k},\mathbf{k'}} \delta_{cc'} \delta_{vv'} \nonumber \\
 = \bar{E}^{\rm QP}_{A,\rm{cond}} - \bar{E}^{\rm QP}_{A,\rm{val}}    
 \end{align}
 is the independent particle transition part of the exciton total energy, or in other words, the exciton total energy minus the exciton binding energy $\Omega_A - \langle A | K^{\rm eh} | A \rangle$.
 
\begin{figure*}
    \centering
    \includegraphics[width=\linewidth]{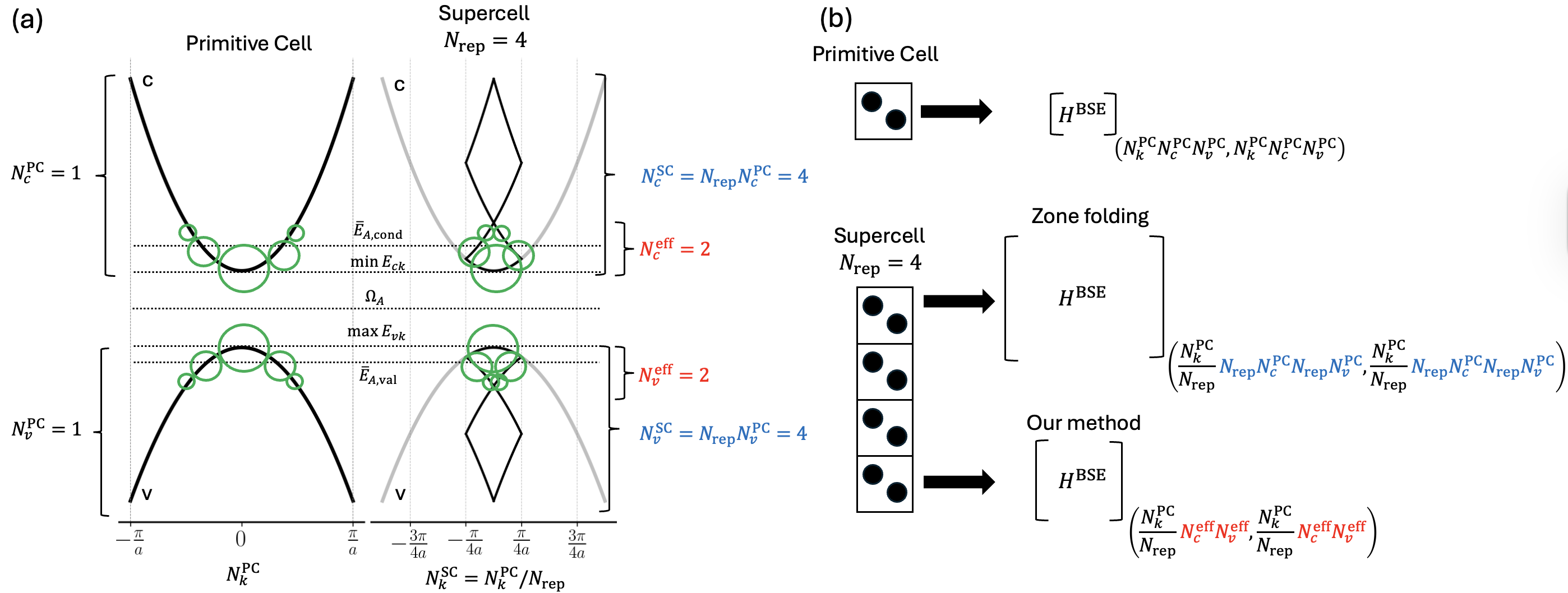}
    \caption{(a) Diagram showing zone folding of the bandstructure and exciton coefficients. We show a 1D example of bands being zone folded in a BZ of a SC four times the size of the PC. Circle radius is proportional to $|A_{cv {\mathbf{k}}}|$. For the PC calculation, only one valence and one conduction band are necessary to converge this exciton wavefunction, while for the SC, two valence and two conduction bands are necessary to represent this exciton, but not four as would be expected by a zone-folding analysis. WASPEs $\bar{E}_{A,\rm{cond(val)}}$ are given by Eq. \ref{eq:energy_window_eq} and give an estimate of the energy scale for the transitions that need to be included. (b) Diagram showing the replication of the PC to create the SC and how this affects the BSE Hamiltonian size. Our method provides a BSE Hamiltonian smaller than the one obtained by zone folding.}
    \label{fig:diagram_zone_folding}
\end{figure*}

 In Fig. \ref{fig:diagram_zone_folding}, we illustrate the basic idea of the method. An exciton is composed of a linear combination of transitions from valence to conduction states and the energy width of the states that compose that exciton is smaller than the whole width of the band \cite{Sun2022, DadkhahPRB2024}, when the transitions that compose this exciton are localized in the Brillouin Zone. In this case, when zone folding, this band will be split into several bands, and some of those may not participate in the composition of that exciton. Therefore, one needs to include in the SC BSE Hamiltonian only the relevant zone-folded bands for the excitons of interest. In practice, this means limiting the sums over the exciton coefficients up to the bands that are inside an energy window $[E^{\rm min}, E^{\rm max}]$. We estimate the exciton energy from PC calculations as
\begin{equation}
    \Omega^{\rm part}_A = 
    \sum_{\substack{
            cv\mathbf{k}, \\
        c'v' \mathbf{k}' \\
    }} 
    \left(A^{\rm part}_{c'v' \mathbf{k}'}\right)^* A^{\rm part}_{cv \mathbf{k}} 
    \langle c'v' \mathbf{k}'|H^{\rm{BSE}} | cv \mathbf{k}\rangle 
\label{eq:omega_part}
\end{equation}
where 
\begin{equation}
    A^{\rm part}_{cv \mathbf{k}} =
    \begin{cases} 
        A_{cv \mathbf{k}} /N, & \text{if } E^{\rm QP}_{c\mathbf{k}} < E^{\rm{max}} \text{ and } E^{\rm QP}_{v\mathbf{k}} > E^{\rm{min}} \\
        0, & \text{otherwise}
    \end{cases}
\label{eq:Akcv_part}
\end{equation}
and $N$ is a normalization factor that makes $\sum_{cv\mathbf{k}} |A^{\rm part}_{cv\mathbf{k}}|^2=1 $.

The more the coefficients $A_{cv\mathbf{k}}$ are spread in energy, the more bands will be necessary to build the BSE Hamiltonian, and less computational savings for a given accuracy of the exciton energy can be achieved. Therefore our method can be applied for both Wannier excitons (localized in $\mathbf{k}$) and Frenkel excitons (localized in real space), although one can expect to achieve a greater reduction of the BSE Hamiltonian in the Wannier exciton cases, where transitions are localized in the BZ, such as the ones in Ref. \cite{Alvertis2023PRB}. In the specific case of LiF, the lowest exciton transitions are localized at the $\Gamma$, but our approach can be used within other materials like monolayer MoS$_2$, in which the lowest exciton has energy 1.88 eV and is composed by transitions localized at the K point \cite{Qiu2013}. We highlight that the current method does not make any assumptions about the materials, so it can be applied to any material where the BSE is applicable.

\subsection{Workflow}

\begin{figure}
    \centering
    \includegraphics[width=\linewidth]{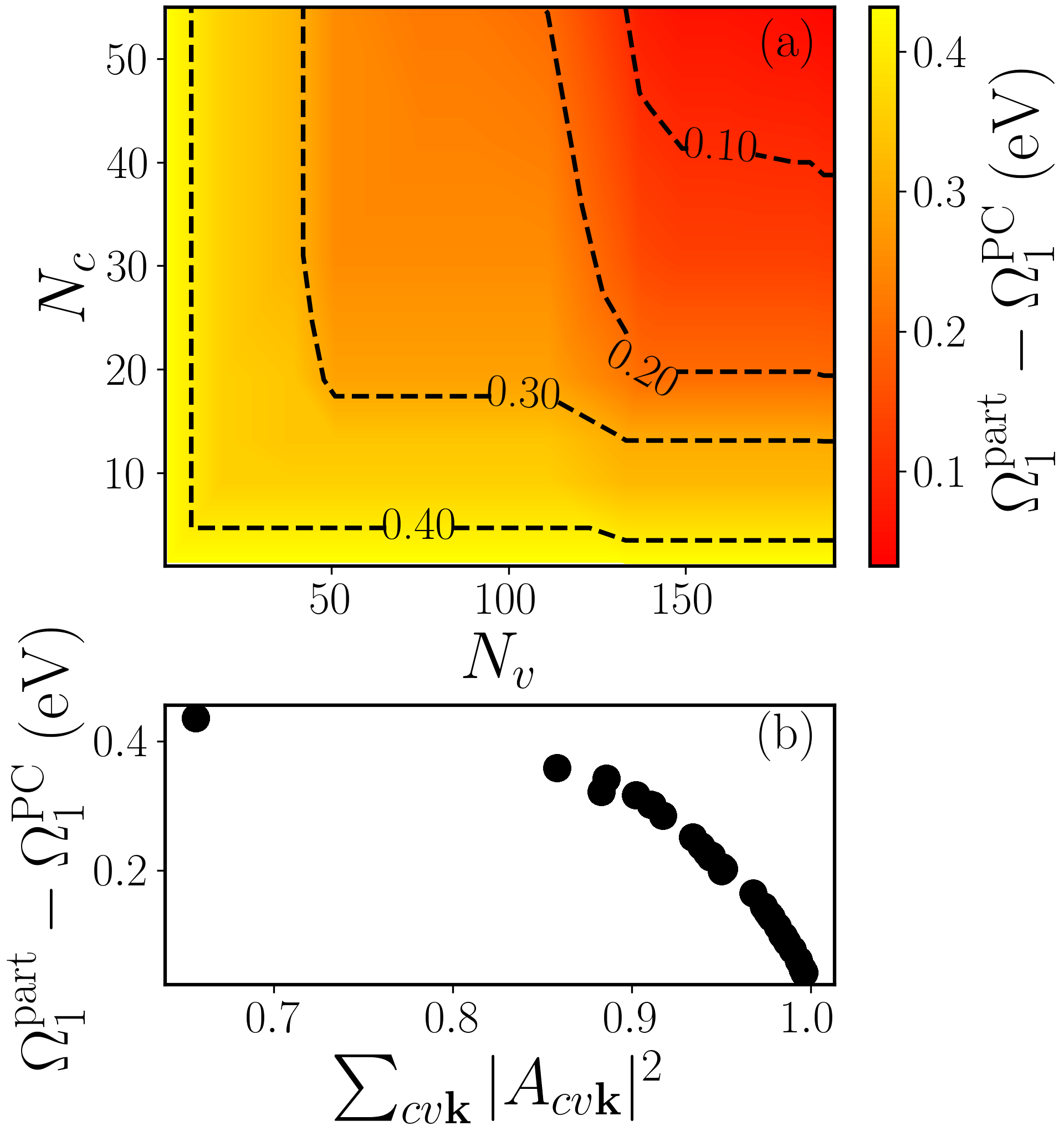}
    \caption{(a) First exciton energy evolution, in comparison to the full PC BSE calculation $\Omega^{\rm{PC}}_1$, as function of the number of valence ($N_v$) and conduction ($N_c$) bands used to build the BSE Hamiltonian. Exciton energy is calculated using Eq. \ref{eq:omega_part}. (b) Correlation of $\Omega^{\rm{part}}- \Omega^{\rm{PC}}_1$ with the partial sum of the exciton coefficients, following a quadratic relation. 
    \label{fig:evolution_exciton_vs_Nc_Nv}}
\end{figure}

To make the choice of the proper number of bands, we recommend the following workflow:

\begin{enumerate}
    \item For a PC with lattice vectors $\boldsymbol{a_1}$, $\boldsymbol{a_2}$, $\boldsymbol{a_3}$, choose a desired SC size with lattice vectors $N_1\boldsymbol{a_1}$, $N_2\boldsymbol{a_2}$, $N_3\boldsymbol{a_3}$ and a desired regular $k$-grid $N_{k,1} \times N_{k,2} \times N_{k,3}$ (chosen based on PC calculations). 
    \item Perform a converged GW/BSE calculation for the PC with a fine $k$-grid $N_1 N_{k,1} \times N_2 N_{k,2} \times N_3 N_{k,3}$. Save the QP energy levels on this fine grid, the eigenvectors file with the $A_{cv\boldsymbol{k}}$ coefficients, and the BSE Hamiltonian matrix elements. All those files can be saved in the absorption step of the BerkeleyGW code \cite{Deslippe2012}, version 4.0 (using the flags  \texttt{dump\_bse} and \texttt{write\_eigenvectors} of the Absorption step). If the BSE Hamiltonian file is not available, as for earlier versions or other codes, the BSE Hamiltonian can be reconstructed using the exciton energies and coefficients through $\langle c'v' \boldsymbol{k}'| H^{\rm{BSE}} | cv \boldsymbol{k}  \rangle = \sum_A \Omega_A A^*_{c'v'\mathbf{k'}} A_{cv\boldsymbol{k}}$.   
    \item  Use the exciton coefficients $A_{cv\mathbf{k}}$ and QP energy levels from the PC to estimate the BSE size for the desired precision of the binding energy in SC calculations. Equation \ref{eq:energy_window_eq} provides an estimate of the necessary energy window, and Eqs. \ref{eq:omega_part} and \ref{eq:Akcv_part} can be used to predict the error in the exciton energy (Fig. \ref{fig:analisys_window_energy}).  In Fig. \ref{fig:evolution_exciton_vs_Nc_Nv}, we show how the partial exciton energy (Eq. \ref{eq:omega_part}) depends on the number of conduction and valence bands. Alternatively, one can plot how much of the exciton weight is preserved inside the chosen energy window, because this quantity has a quadratic relation with the partial exciton energy (Fig. \ref{fig:evolution_exciton_vs_Nc_Nv}(b)). We provide a Jupyter notebook and example data for this analysis \cite{supplemental_material} that is compatible with BerkeleyGW \cite{Deslippe2012}. The Jupyter notebook also has the code necessary to recover the BSE Hamiltonian from the BSE eigenvectors using the equation in the end of step 2.

\end{enumerate}

The present workflow can be applied to any exciton of interest (specified by its index). Our Jupyter notebook runs in just a few minutes in a personal computer, allowing one to explore the convergence of several excitons. In our case of LiF, both excitons are localized at $\Gamma$ and their convergence is pretty similar (Fig. \ref{fig:absorption_PC_vs_SC}). By contrast, in monolayer MoS$_2$, the A exciton is localized at K and the C exciton is composed of transitions at K and around $\Gamma$ \cite{Qiu2013}. In this case, one needs to make the convergence study for each exciton separately and use the larger of the number of bands in order to converge both simultaneously. In general, higher excitons may spread more in energy and in $k$ space, which would require more bands to be converged. These calculations are done by selecting the same number of valence (conduction) bands per k-point, which is what is done in BerkeleyGW as part of the memory layout to enable efficient parallel matrix-vector operations \cite{Deslippe2012}. Alternate schemes in which arbitrary pairs of states are selected could be used to reduce the basis further but would be inconsistent with this layout for computational performance.
Trying to use folding back to a PC BZ in order to exclude transitions between different PC $\mathbf{k}$ points is unlikely to be helpful because cases of interest for a SC calculation are the ones where there is a breaking of symmetry and the mixing of states from different $\mathbf{k}$-points may be important for the optical properties of interest, e.g. brightening of dark excitons, so such an approach could save some computational time but some of the physics may be lost as well. 

\section{Computational Details}
\label{sec:computational_details}

All calculations in this work were performed for LiF in the rocksalt structure with the lattice parameter equal to its experimental value at 300 K, 4.026 {\AA} \cite{BenYahia2012, Srivastava1973, Shadike2021}. We used ONCV scalar-relativistic LDA pseudopotentials with standard precision from pseudodojo \cite{VansettenCompPhysCom2018, HamannPRB2013}. Parameters for DFT calculations are: $E_{\rm{cut}} = 80 \ \rm{Ry}$ and a regular $\Gamma$-centered $4 \times 4 \times 4$ $k$-grid for PC calculations. For GW/BSE calculations: coarse and fine grids are $4 \times 4 \times 4$, and no interpolation from coarse to fine grids is done. The cutoff energy to build the dielectric matrix is 20 Ry. We performed G$_0$W$_0$ calculations within the Generalized Plasmon Pole \cite{Deslippe2012, Hybertsen1986} using the stochastic pseudobands method \cite{AltmanPRL2024} with an accumulation window of 2\%, 2 stochastic pseudobands per energy subspace, and generating empty states with energy up to the DFT cutoff (80 Ry). The pseudobands method is 
useful for large systems and can result in a two orders-of-magnitude speed-up in GW calculations on supercells of this size \cite{AltmanPRL2024}, but our method can be used with or without pseudobands.

The PC BSE Hamiltonian used for the calculation results is built with 5 valence and 10 conduction bands. Only 3 valence and 1 conduction bands are necessary to converge the first optical absorption peaks, and we present later timing comparisons from this minimal set of bands, but we include more bands in other calculations to allow a more detailed exploration of a wider range in the spectrum. Optical absorption plots reported in this paper were calculated with light polarization parallel to the direction $-\hat{x}+\hat{y}$, with the momentum operator \cite{Deslippe2012, Rohlfing2000} (which does not include non-local effects) for simplicity, as our objective in this work is only to study the convergence. 

The SC is composed of $4\times 4\times 4$ PCs and has 128 atoms. For a faithful comparison, both PC and SC have LiF in the rocksalt structure, and our objective in this example is to make SC calculations reproduce results from the PC case. Therefore, both coarse and fine SC $k$-grids include only the $\Gamma$-point, satisfying the $k$-grid condition for exact correspondence. Except for the number of conduction and valence states to build the BSE Hamiltonian, all other GW/BSE and DFT parameters are the same as in the PC case. 

DFT calculations were performed with the Quantum ESPRESSO code \cite{Giannozzi2009, Giannozzi2017, Giannozzi2020} (version 7.2) and GW/BSE calculations were performed with the BerkeleyGW code \cite{Deslippe2012, Rohlfing2000, Hybertsen1986} (version 4.0), but this approach is general and can be used by other codes that solve the BSE.

\section{Results}
\label{sec:results}

\subsection{GW results}

\begin{figure}
    \centering
    \includegraphics[width=\linewidth]{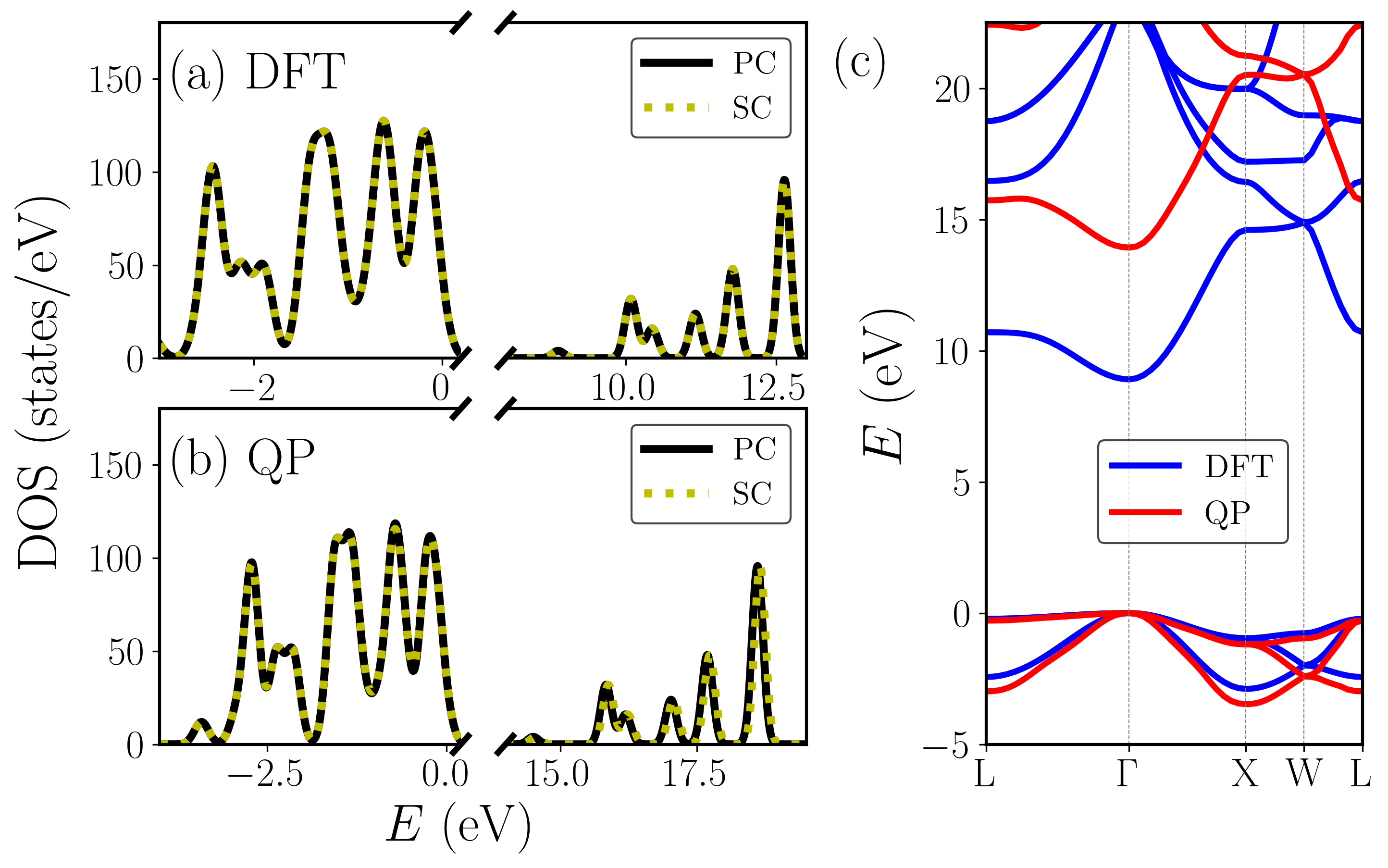}
    \caption{Density of States (DOS) at DFT (a) and QP (b) levels for rocksalt LiF calculations in a 2-atom PC (black solid lines) and a 128-atom SC (green dashed lines). Direct bandgap energies at the $\Gamma$-point are reported in Table \ref{tab:comparison_energy_peaks_PC_SC}. For DOS calculation, a Gaussian broadening of 0.1 eV was used.
    (c) DFT and QP bandstructure of LiF, from PC. In all panels, we set the maximum of the valence band to be zero.}
    \label{fig:DOS_primitive_vs_supercelll}
\end{figure}

To test this approach, we show results for rocksalt LiF. For comparison, we show in Fig. \ref{fig:DOS_primitive_vs_supercelll} the Density Of States (DOS) for the SC and PC calculations at DFT and QP levels. The curves are almost identical between SC and PC, for both DFT and QP DOS. For the PC case, we find a GW bandgap at $\Gamma$ equal to 14.502 eV, while for SC, the bandgap is 14.550 (Table \ref{tab:comparison_energy_peaks_PC_SC}). The bandgap difference at the GW level is 48 meV, which is 0.3\% of the PC GW gap. The difference is due to the approximations used in the stochastic pseudobands method \cite{AltmanPRL2024}. By contrast, the bandgap difference at the DFT level is 3 meV, which is only due to subtle differences in numerical convergence.  Because of the offset in the bandgap, we will compare the convergence of the exciton binding energies of PC and SC calculations, rather than the absolute exciton energies. We define the binding energy as the QP bandgap (which occurs at ${\bf k} = \Gamma$) minus the exciton energy, which is a positive number.

\subsection{How to choose the BSE size for SC calculations}

\begin{figure} 
    \centering
    \includegraphics[width=\linewidth]{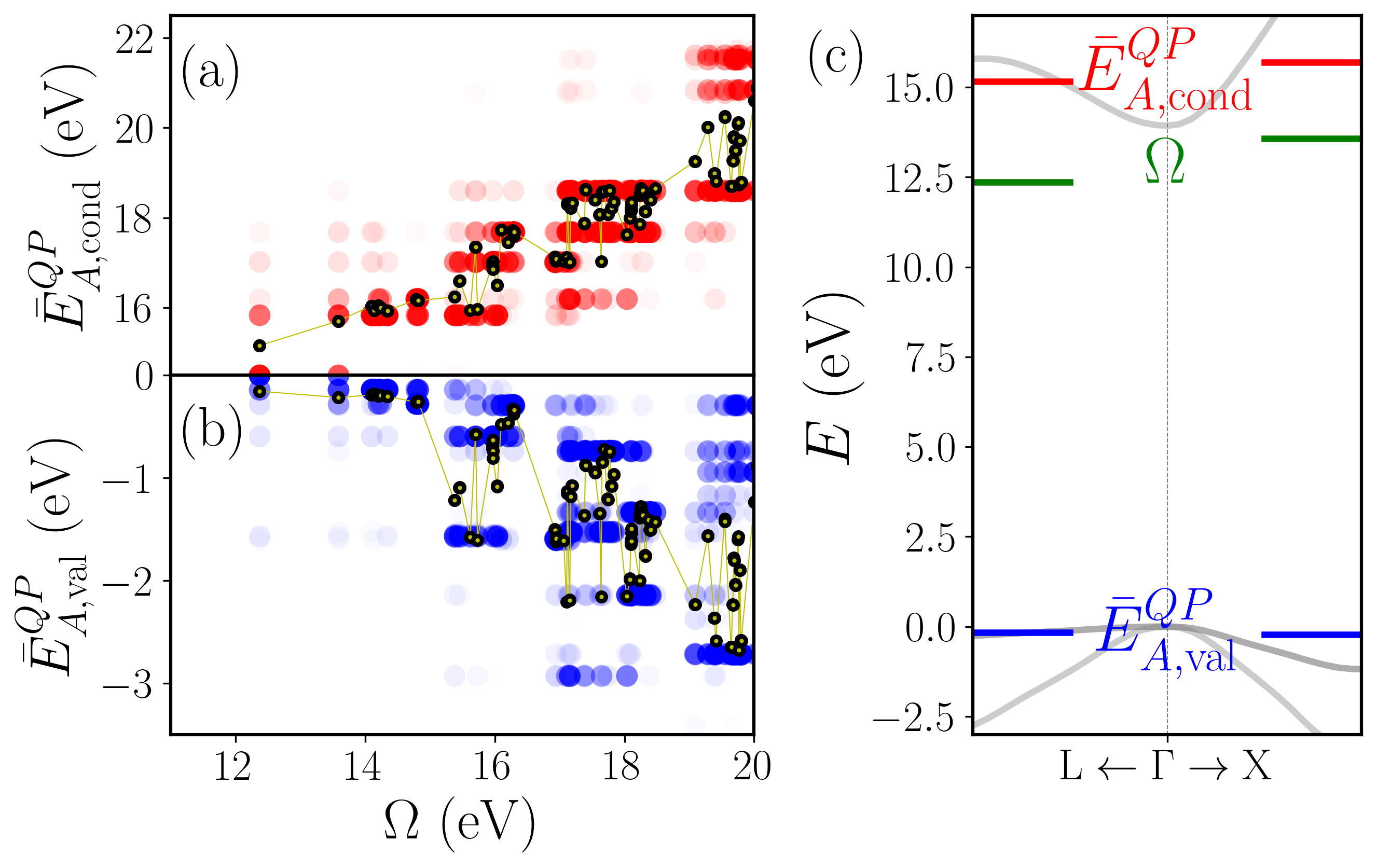}
    \caption{Energy analysis for excitons with energy ranging 11.7 to 20.0 eV. (a-b) Small dots are the conduction (a) and valence (b) WASPEs (Eq. \ref{eq:energy_window_eq}) for excitons with excitation energy $\Omega$. Circles are located at the energies of the states that compose the exciton, and the color intensity is proportional to $|A_{cv{\bf k}}|^2$, showing the distribution in energy of the states that compose this exciton.
    (c) QP bandstructure for PC around $\Gamma$ point. Horizontal green lines are the exciton energies for the first (left) and second (right) bright excitons. Red (blue) lines are the conduction (valence) WASPE corresponding to these excitons.}
    \label{fig:analisys_window_energy}
\end{figure}

Then we calculate the exciton energies for PC LiF using a $4\times 4 \times 4$ $k$-grid, five valence bands, and ten conduction bands to build the BSE Hamiltonian. Due to the Brillouin zone folding, just the $\Gamma$ point is necessary for SC BSE calculations. Zone folding leads to 320 valence and 640 conduction bands. To choose a reasonable number of bands, we show $\bar{E}^{\rm{QP}}_{A,\rm{val}}$ and $\bar{E}^{\rm{QP}}_{A,\rm{cond}}$ for several excitons in Fig. \ref{fig:analisys_window_energy}. We note that they generally increase with exciton energy. Valence band energies vary less than conduction band energies, as the bandwidth of the conduction band for LiF is larger than for the valence band (Fig. \ref{fig:DOS_primitive_vs_supercelll}). We also plot the exciton coefficient distribution for each exciton (corresponding to the color intensity in the figure). The exciton coefficients go out to a few eV beyond the WASPEs, which indicates that the energy window needs to include those transitions.

\begin{figure*}
    \centering
    \includegraphics[width=0.8\linewidth]
    {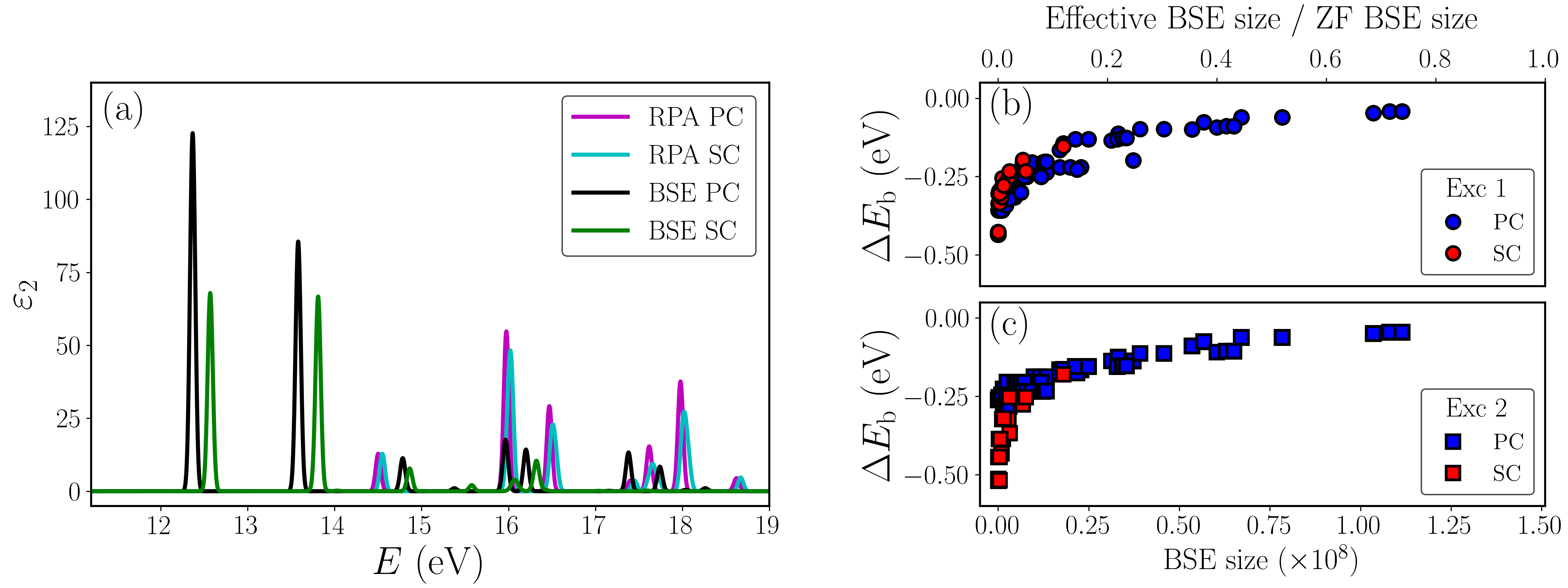}
    \caption{(a) Optical absorption for PC and SC cases calculated at BSE and at QP-Random Phase Approximation (RPA) levels. For the SC, we used 137 valence and 31 conduction states to build the BSE Hamiltonian. Absorption peaks from SC calculations are blueshifted by $\sim 0.15$ eV. (b) Convergence of first exciton binding energy using PC results and Eq. \ref{eq:omega_part} (see text) in blue, and using SC in red, varying the BSE size (number of matrix elements). The agreement between the two data sets is excellent. (c) Same as (b) but for the second bright exciton. In the top $x$-axis, the zone-folding (ZF) BSE size is $((3\times64)\times(1\times64))^2 \sim1.5 \times 10^8$. The reference binding energies for the PC at full size are in Table \ref{tab:comparison_energy_peaks_PC_SC}.}
    \label{fig:absorption_PC_vs_SC}
\end{figure*}

Next, we estimate SC results from PC results focusing on the first and second absorption peaks (Fig. \ref{fig:absorption_PC_vs_SC}(a)). In Fig. \ref{fig:absorption_PC_vs_SC}(b) and (c), we plot the expected binding energy variation with respect to PC results for the first and second absorption peaks, respectively. The expected binding energy was given by the expected exciton energy from Eq. \ref{eq:omega_part}, subtracted from the QP energy gap. In our code (provided at \cite{supplemental_material}), we count how many valence and conduction states $N_c$, $N_v$ are present in a given energy window $[E^{\rm{min}}, E^{\rm{max}}]$. Then we estimate the value of $\Omega^{\rm part}_A$ for the BSE size $(N_c N_v N_k^2)^2$. The expected binding energy increases roughly monotonically with the BSE size until it reaches the PC results when the BSE size is equal to the size of a BSE Hamiltonian obtained by zone folding. The reference binding energies for the PC can be found in Table \ref{tab:comparison_energy_peaks_PC_SC}. Beyond the lowest energy excitons, the energy window to ensure convergence increases, so more bands are necessary to build the BSE Hamiltonian as shown in Fig. \ref{fig:analisys_window_energy}.

To test the quality of the predictions made by PC calculations, we performed SC calculations with different BSE Hamiltonian sizes, and the agreement between the two data sets is excellent (Fig. \ref{fig:absorption_PC_vs_SC} (b) and (c)). This shows the efficiency of our analysis in predicting results on BSE SC calculations based on the results of PC. In our final result for SC calculations, the BSE is composed of 137 valence bands and 31 conduction bands with 1 $k$-point (just $\Gamma$), which means that $(1\times137\times31)^2\sim 18\times 10^6$ matrix elements are necessary. By zone-folding 3 valence and 1 conduction bands of a $k$-grid with 64 $k$-points of the PC into the SC Brillouin zone, the SC calculation would need 3$\times$64 valence, 64 conduction bands and 1 $k$-point, which means that $(3\times64\times64\times1)^2\sim 151 \times 10^6$ matrix elements would be necessary. Therefore, our approach requires only 12\% of the matrix elements one would expect to need from only using zone folding. The error is 150 meV for the first exciton, which is 1.2\% of the first excitation energy value. In Table \ref{tab:timing_PC_vs_SC}, we show some time estimation comparisons, showing how our approach can save around a factor of 10 in the time for SC BSE calculations, given the reduction in the number of matrix elements by 12\%. Next, we show the optical absorption spectra for PC and SC cases in Fig. \ref{fig:absorption_PC_vs_SC} (a). We observed that for energies up to 17 eV, BSE absorption peaks for SC are blueshifted by 0.15 eV in relation to PC results (Table \ref{tab:comparison_energy_peaks_PC_SC}), while no absorption peak is present for energies above 17 eV in the SC BSE spectrum with the limited number of bands used.

\begin{table*}
    \centering
    \begin{tabular}{|c|c|c|c|c|}
    \hline
        
          & \multicolumn{2}{c|}{This work} & \multicolumn{2}{c|}{Other works}  \\ \hline
        Energies (eV) & PC  & SC & Theory  & Experiment  \\
        \hline \hline
        DFT Gap at $\Gamma$ & 8.8723& 8.8756 & 8.91 \cite{WangPRB2003} &\\
        QP Gap at $\Gamma$  & 14.502& 14.550 & 14.4 \cite{Rohlfing2000}, 14.3 \cite{WangPRB2003}, 14.5 \cite{Dai2024PRB} & $14.2\pm0.2$ \cite{PiacentiniPRB1976}\\
        $\Omega_1$          & 12.368& 12.570& 12.8 \cite{Rohlfing2000}, 12.7 \cite{WangPRB2003}, 12.82 \cite{Dai2024PRB} & 12.5 \cite{Roessler1967}\\
        $\Omega_2$          & 13.581& 13.811& & \\
        $E_{\rm b1}$          & 2.134& 1.980&  1.6 \cite{Rohlfing2000}, 1.88 \cite{Dai2024PRB} & \\
        $E_{\rm b2}$          & 0.921& 0.739& & \\
    \hline
    \end{tabular}
    \caption{Bandgap at $\Gamma$ at DFT and QP levels, exciton energies for first ($\Omega_1$) and second ($\Omega_2$) absorption peaks (Fig. \ref{fig:absorption_PC_vs_SC}), and respective binding energies $E_{\rm b}$ for calculations in SC and PC. The binding energy is defined as $E^{\rm{QP}}_{\rm{gap}}-\Omega$.}
    \label{tab:comparison_energy_peaks_PC_SC}
\end{table*}

\subsection{Relation of excitonic properties with BSE size}

\begin{figure}
    \centering
    \includegraphics[width=\linewidth]{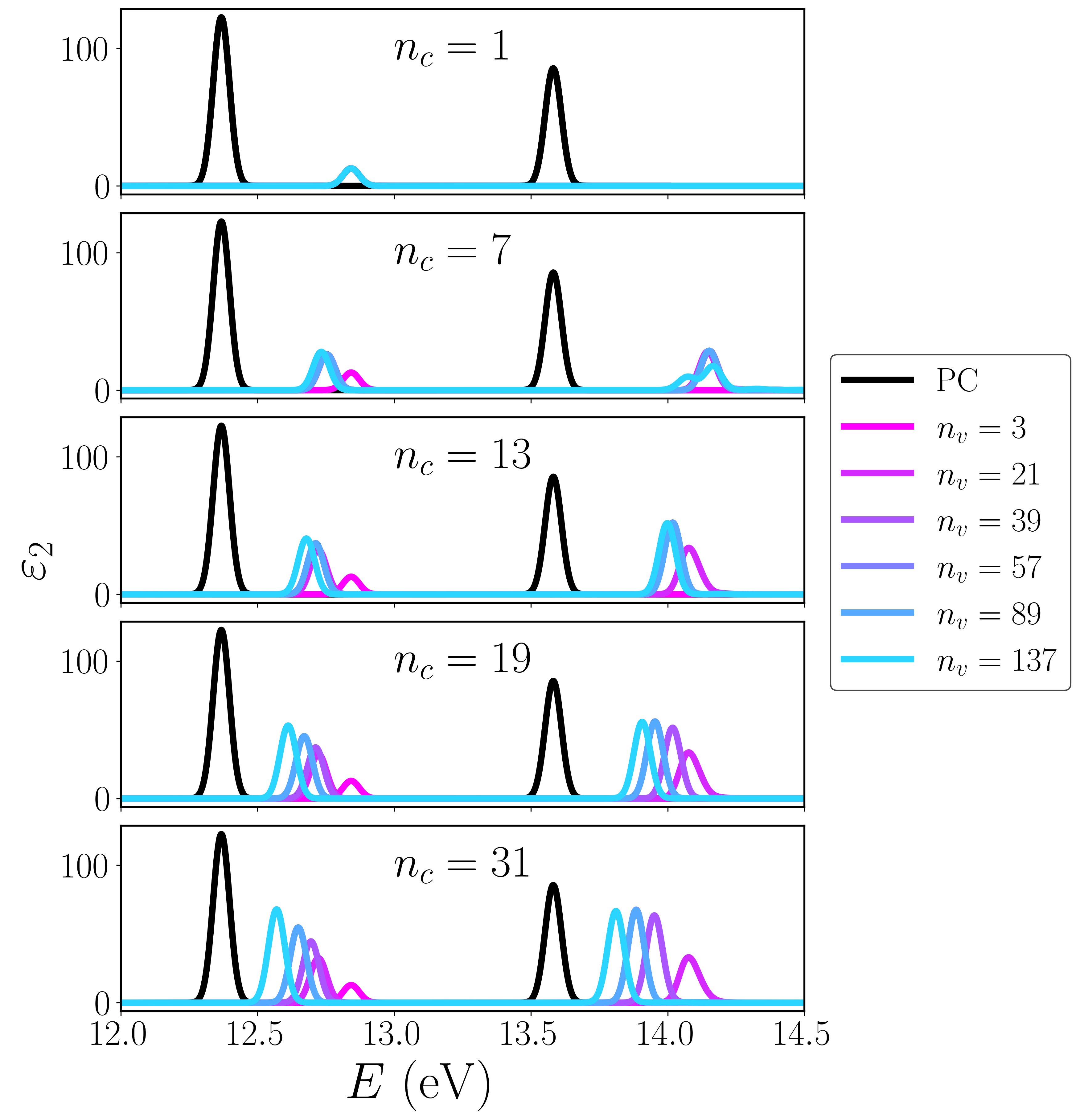}
    \caption{Optical absorption varying the number of conduction and valence bands in the BSE Hamiltonian. The black curve is the PC result, for the fixed size $N_v= 5$, $N_c = 10$ as described in Section \ref{sec:computational_details}, while the colored curves are SC results varying the BSE Hamiltonian size.}
    \label{fig:absorption_LiF_vary_kernel_size}
\end{figure}

In Fig. \ref{fig:absorption_LiF_vary_kernel_size}, we show the evolution of the absorption spectra when varying the size of the SC BSE Hamiltonian. SC peaks are blueshifted in comparison to PC results, given the variational nature of the BSE diagonalization and limited number of bands. The peak intensities are underestimated as well. As the BSE size increases, the exciton energies decrease, and also the peaks' intensities increase toward PC results. 

\begin{figure}
    \centering
    \includegraphics[width=\linewidth]{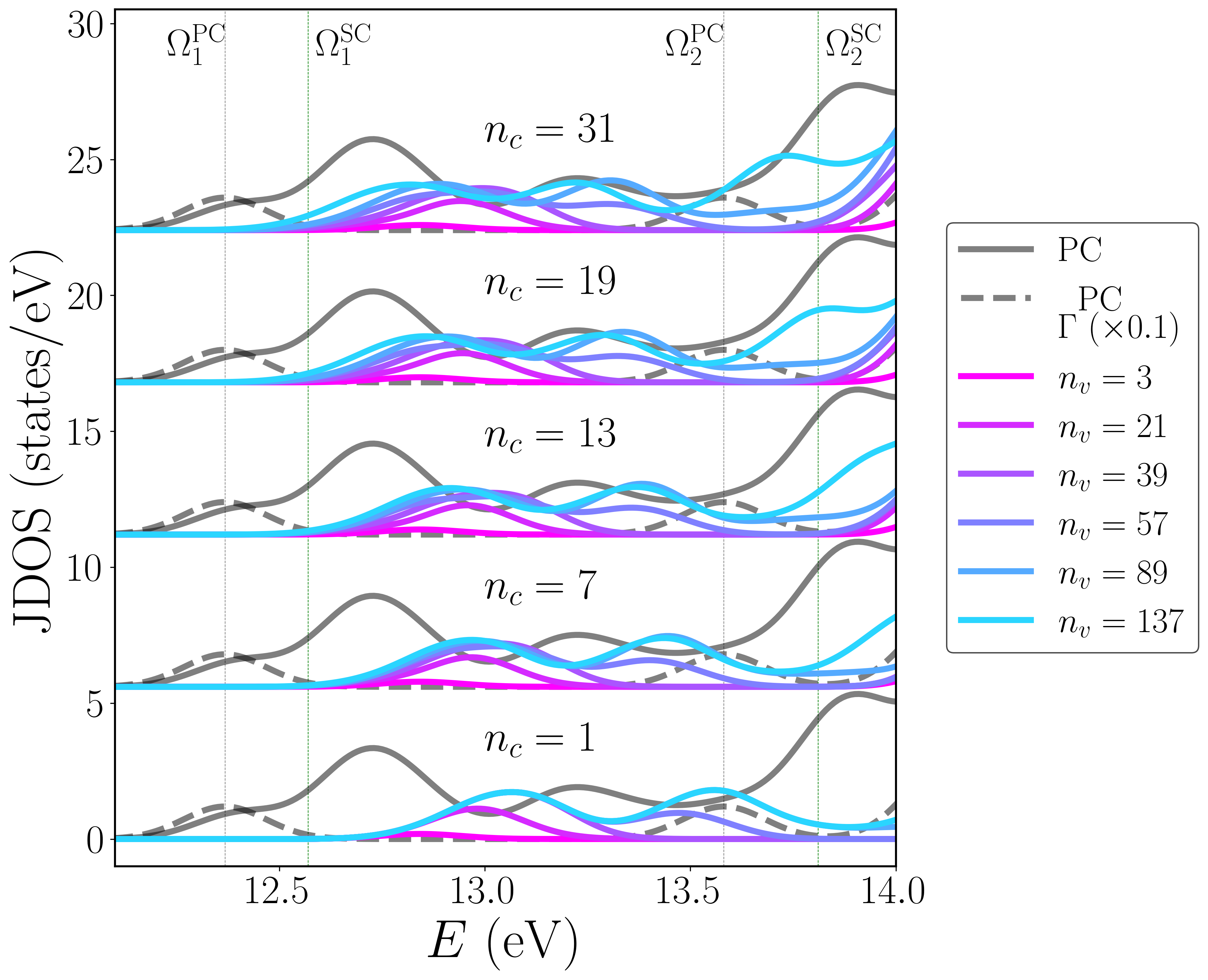}
    \caption{Joint Density of States (Exciton Density of States) for PC calculations (solid and dashed gray lines) compared with SC calculations with different numbers of conduction and valence states to build the BSE Hamiltonian. We show two PC calculations, one where just excitons with zero center-of-mass momentum are included in the calculation (dashed gray line) and the other where all excitons with finite momentum $Q$ from a regular grid $4\times 4 \times 4$ are included (solid gray line). Due to zone folding, a BSE calculation for the SC results in excitons corresponding to finite momentum from the PC. The more bands used to build the BSE Hamiltonian, the closer the SC exciton JDOS gets to the PC exciton JDOS. Vertical lines are the energies of the first and second bright excitons (Table \ref{tab:comparison_energy_peaks_PC_SC}).}
    \label{fig:JDOS_evolution_BSE_size}
\end{figure}

One important aspect to be considered is that due to zone-folding, finite-momentum excitons of PC may be folded into the SC results. To illustrate this, we show in Fig. \ref{fig:JDOS_evolution_BSE_size} the Joint Density Of States (JDOS) of SC calculations and compare it with the PC case. For a fair comparison, the PC BSE results must include finite-momentum excitons that will be folded into the SC Brillouin Zone. Our finite momentum results include excitons with momentum $Q$ \cite{QiuPRL2015}, where $Q$ is a $q$-vector of a regular $4 \times 4 \times 4$ grid. As we increase the number of bands to build the BSE Hamiltonian, the JDOS gets closer to PC results with finite momentum excitons. PC JDOS with just excitons at $Q=0$ ($\Gamma$ point) is zero in the region between 12.5 and 13.0 eV, but in our SC result the JDOS is finite in agreement with PC JDOS using excitons with finite momentum. Finite momentum excitons are dark as photons carry no momentum, and here we are comparing PC and SC calculations for the same RS structure, so the optical absorption for the two cases is almost identical. However, if one studies cases where there is a symmetry-breaking in the SC case (e.g. a point defect), which can brighten those dark excitons, it is necessary to use SC parameters corresponding to a well-converged PC JDOS as well as absorption spectrum. In Fig. \ref{fig:JDOS_evolution_BSE_size}, we show how the JDOS of the SC evolves when increasing the BSE size, and our choice with 137 valence and 31 conduction bands agrees better with the PC JDOS with excitons with finite $Q$. Full convergence of the JDOS to the same level as the absorption spectrum would still require more bands, as the dark and $Q \ne 0$ excitons evidently are harder to converge.

\section{V$_k$-like defect in LiF}

\label{sec:Vk_center_results}
\begin{figure}
    \centering
    \includegraphics[width=1\linewidth]{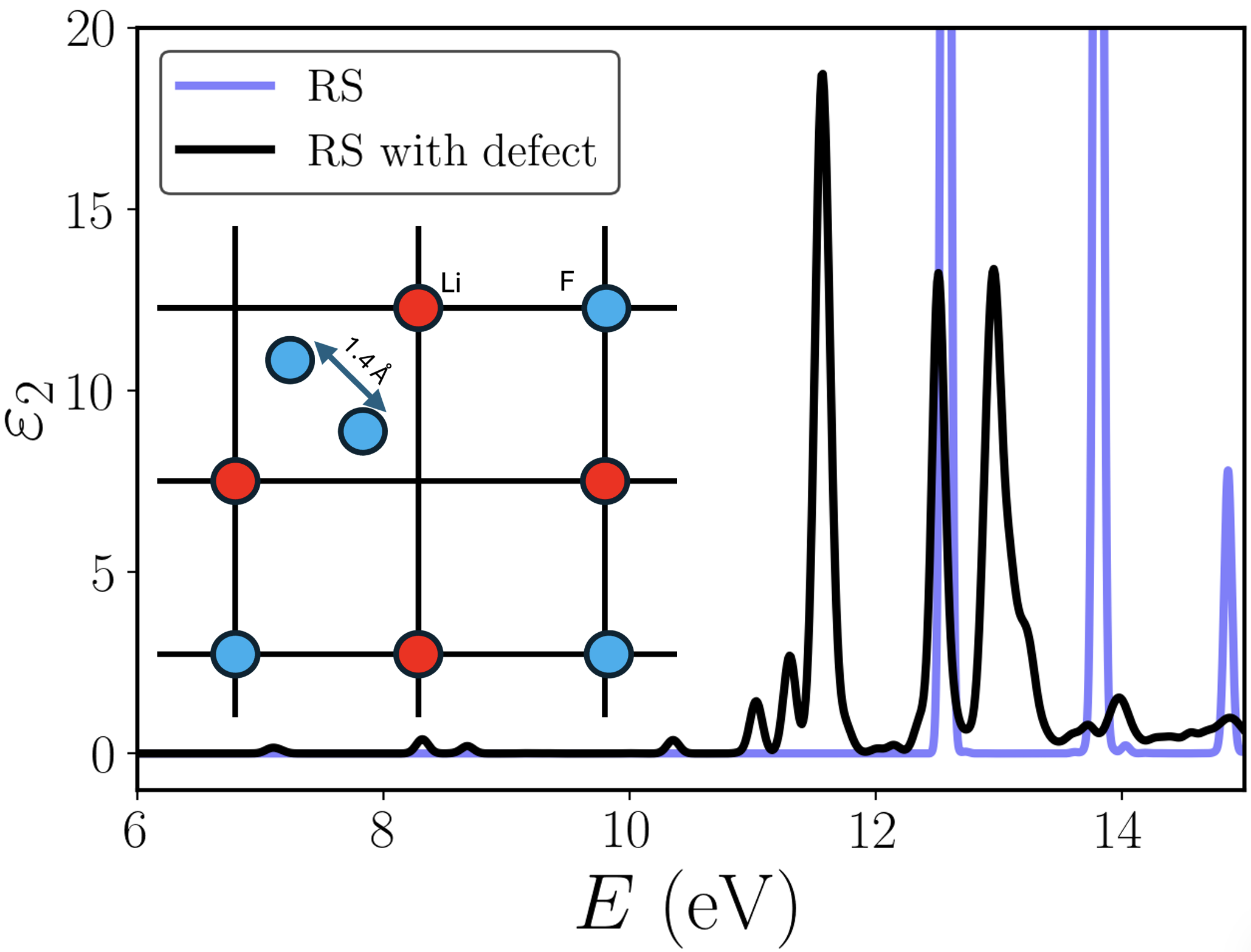}
    \caption{Optical absorption for rocksalt (RS) LiF and a V$_k$-like defect, both calculated in a 4$\times$4$\times$4 SC. Inset: Structure of the defect in which 2 fluorine atoms are moved towards each other in the [110] direction. }
    \label{fig:vk_like_abs}
\end{figure}

As a proof of concept we performed a GW/BSE calculation for a V$_k$-center-like defect in LiF \cite{Pandey1988}, and our absorption spectra results are shown in Fig. \ref{fig:vk_like_abs}. V$_k$ centers in alkali halides are defects where the halide atoms get closer to each other, similar to what is shown in Fig. \ref{fig:vk_like_abs}. To represent well the V$_k$ center one would need to include the spin degree of freedom and relax the excited state \cite{delgrande2025arxiv} for the defect exciton, which we did not do as it is out of the scope of this work. This result illustrates how defect calculations can be performed using our scheme. It uses the same parameters of our SC BSE calculations reported in Table \ref{tab:comparison_energy_peaks_PC_SC}, notably the number of valence and conduction bands used in the BSE, which was provided by our method. Two fluorine atoms are moved towards each other in the [110] direction and the distance between them is 1.4 {\AA}. This is a somewhat arbitrary displacement to break the symmetry which can be used for further excited-state relaxations\cite{delgrande2025arxiv}. In our BSE results we can observe that the two lowest peaks of the case without defect are still present, and the second lowest peak is redshifted. In addition to that, new peaks between 6 and 12 eV appear, which are transitions including mid gap defect states.

\section{Conclusions}
\label{sec:conclusions}

We presented in this paper an effective scheme to set the BSE Hamiltonian size for SC calculations based on PC results without the need for new convergence studies, which we demonstrated with detailed analysis of the case of rocksalt LiF. The scheme is applicable to exact SCs or cases of slightly perturbed atomic positions, in which case analysis of finite-momentum excitons in the PC is likely also needed, to converge the unperturbed JDOS as well as the absorption spectrum. We introduced the Weighted Average Single-Particle Energies (WASPEs) for conduction and valence of each exciton, to indicate the scale of energies that need to be included in the SC BSE Hamiltonian, and we further analyzed the energy window of states participating significantly, which may extend for another 1-2 eV, to determine the states needed. We found a reduction of an order of magnitude in the number of matrix elements needed, leading to an order of magnitude reduction in computational expense for BSE. Our analysis shows that the kernel matrix calculation is the key concern since its expense rises more rapidly with supercell size than the absorption calculation in which the matrix is diagonalized. Our approach for determining the number of states needed in BSE also allows one to choose for which energy levels GW calculations are necessary, avoiding unnecessary calculations and making SC GW/BSE calculations more accessible. Then we showed, as a proof of concept, BSE calculations on a V$_k$-like defect in LiF, demonstrating the applicability of our method. We provide a Jupyter notebook and example data \cite{supplemental_material} with the analysis from PC results that can be used for materials in general. This method may be applied to study the optical properties of SC models such as surfaces, defects, and polarons in diverse materials, speeding up \textit{ab initio} results and decreasing RAM demands of exciton calculations substantially, and enabling studies of more complex quantum effects in materials, and can be used within any code that builds the BSE Hamiltonian as described in the text. 

\section*{Acknowledgments}

R.R.D.G. and D.A.S. were supported by the U.S. National Science Foundation under Grant No. DMR-2144317. Computational resources were provided by the National Energy Research Scientific Computing Center (NERSC), a U.S. Department of Energy Office of Science User Facility operated under Contract No. DE-AC02-05CH11231, using NERSC award
FES-ERCAP0028527; the Texas Advanced Computing Center (TACC) at the University of Texas at Austin (http://www.tacc.utexas.edu); and the Pinnacles cluster (NSF MRI, \#2019144) at the Cyberinfrastructure and Research Technologies (CIRT) at the University of California, Merced. We acknowledge Elsa Vazquez for helpful comments on the manuscript.




%

\end{document}